# Computational screening of amorphous lithium trivalent-metal chloride solid-state electrolytes†

Xiaolin Liu, Shuang (Sarah) Xia and De-en Jiang*

Department of Chemical and Biomolecular Engineering, Vanderbilt University, Nashville, Tennessee 37235, USA.

E-mail: de-en.jiang@vanderbilt.edu

Amorphous solid-state electrolytes (SSEs) are attractive candidates for safe, high-energy-density all-solid-state batteries. The recent discovery that amorphous $LiTaCl_6$ and $LiNbCl_6$ reach room-temperature ionic conductivities above 10 mS cm$^{-1}$ with activation energies below 0.2 eV has spurred a broad search for new amorphous halide Li conductors. Here we report a computational screening of amorphous lithium trivalent-metal chlorides (Li–M(III)–Cl) using molecular dynamics driven by a universal machine-learning force field. For the sixteen Li–M(III)–Cl compositions of 15 rare-earth elements and Indium, all phases exhibit fast room-temperature $Li^+$ diffusion, with diffusivities of roughly 2–5 × 10$^{-7}$ cm$^2$ s$^{-1}$. After scaling the conductivity vs diffusivity relation to an experimental $LiTaCl_6$ reference, the estimated room-temperature conductivities range from 7.7 mS cm$^{-1}$ for amorphous $Li_3InCl_6$ to 14 mS cm$^{-1}$ for amorphous $Li_3ScCl_6$. With a low activation energy of 0.18 eV, amorphous $Li_3ScCl_6$ is predicted to be among the most conductive amorphous halide electrolytes. To rationalize the composition-dependent transport, we identify two descriptors: the $Li^+$ vibrational density of states (VDOS) center, which correlates inversely with diffusivity, and the average number of correlated bodies in string-like $Li^+$ motion. Compositions in which $Li^+$ has a lower VDOS center and migrates more by individual hopping rather than by concerted motion exhibit higher Li-ion diffusivities. These results suggest rich Li-ion transport characteristics in amorphous halide SSEs.

† Electronic supplementary information (ESI) available: compositions, M–Cl coordination numbers and equilibrium densities (Table S1); UMA+D3 vs. PBE+D3 energy and force parity plots (Figs. S1–S3, S6); radial distribution functions, $Li^+$ diffusivity and Arrhenius analysis for amorphous $LiTaCl_6$ (Figs. S4, S5); Arrhenius analysis for amorphous $Li_3InCl_6$ (Fig. S7); RDFs of all screened phases (Fig. S8); sensitivity of $N_{avg}$ to $\Delta t$ (Fig. S9); stability of all screened phases (Fig. S10). See DOI: https://doi.org/10.1039/x0xx00000x

## 1 Introduction

All-solid-state batteries (ASSBs) promise improved safety and energy density relative to conventional lithium-ion batteries that rely on flammable liquid electrolytes, but their performance hinges on solid-state electrolytes (SSEs) with high ionic conductivity.[1–3] Progress has been made across several SSE families, including oxide garnets such as $Li_7La_3Zr_2O_{12}$ (LLZO),[4] sulfides such as $Li_{10}GeP_2S_{12}$ (LGPS),[5] the argyrodite $Li_6PS_5Cl$,[6] and crystalline halides such as $Li_3InCl_6$[7] and $Li_3ScCl_6$.[8] Among crystalline SSEs, oxides such as LLZO offer wide electrochemical windows but are brittle and form poor solid–solid contact with electrodes;[9] sulfides offer the highest conductivities (10–20 mS $cm^{-1}$)[5,10] but suffer from narrow electrochemical windows[11] and air sensitivity;[12] and halides combine good oxidative stability and processability with more modest conductivities.[13–15]

Strikingly, amorphous $LiTaCl_6$[16] and $LiNbCl_6$[17,18] exhibit room-temperature $Li^+$ conductivities exceeding 10 mS $cm^{-1}$ and activation energies below 0.2 eV, with amorphous $LiNbCl_6$ reaching 13.5 mS $cm^{-1}$ and an activation energy of 0.154 eV which is among the lowest reported for any lithium conductor. These results raise a natural question: how general is superionic transport in amorphous chlorides, and can it be extended across a broader chemical space? A particularly attractive arena is the family of trivalent-metal chlorides, Li–M(III)–Cl, in which M is a trivalent cation combined with LiCl to form ternary lithium halides such as $Li_3MCl_6$ and $Li_6MCl_9$. Several features make this family compelling for amorphous superionic design. First, the trivalent oxidation state demands a high lithium content for charge neutrality: three Li per M in $Li_3MCl_6$, versus a single Li per M in the pentavalent $LiTaCl_6$ and $LiNbCl_6$, supplying a larger reservoir of mobile carriers. Second, the M(III) chlorides display unusually rich coordination chemistry: compact cations favor octahedral $MCl_6^{3-}$ units, whereas larger cations such as $La^{3+}$ adopt higher coordination numbers, forming nine-coordinate $LaCl_9^{6-}$ polyhedra and lithium-rich stoichiometries such as $Li_6LaCl_9$. Third, the trivalent metals span a wide and continuously tunable range of ionic radii from compact $Sc^{3+}$ and $In^{3+}$ through $Y^{3+}$ to the large early lanthanides, allowing the influence of cation size on the anion framework and on $Li^+$ transport to be probed systematically within a single chemical family.[19] This structural diversity has already been exploited synthetically: ball-milling $SmCl_3$ with LiCl yields a crystalline framework with enlarged one-dimensional $Li^+$ channels enclosed by $SmCl_9^{6-}$ tricapped trigonal prisms,[20] while a related $LaCl_3{\cdot}TaCl_5{\cdot}LiCl$ composite opens three-dimensional, interconnected $Li^+$ pathways and pushes the ionic conductivity still higher.[21] Finally, pentavalent $LiTaCl_6$ and $LiNbCl_6$ are reduced at the Li interface and require a protective buffer layer,[16] and the +3 oxidation state offers a plausible route to lithium-metal compatibility.

Exhaustively characterizing transport across so many compositions by ab initio molecular dynamics is computationally prohibitive. Machine-learning force fields (MLFFs) trained on the fly have enabled accurate simulations of the transport and elastic properties of amorphous lithium superionic conductors,[18,22–24] but each must be trained for a specific system. The recent emergence of universal, or foundational, MLFFs such as MACE,[25] UMA,[26] and GRACE[27] now enables accurate and transferable simulations across diverse chemistries without system-specific training, making a broad screen feasible. Here we use the UMA potential with D3 dispersion, validated against PBE+D3 density-functional theory, to screen the room-temperature $Li^+$ transport of amorphous Li–M(III)–Cl compositions spanning much of the rare-earth series together with indium, as some of their crystalline phases such as $Li_3ScCl_6$, $Li_3YCl_6$, $Li_3ErCl_6$, and $Li_3InCl_6$ have been found to be good SSEs.[21] We aim to identify promising amorphous Li–M(III)–Cl systems and compare with their crystalline counterparts as well as to establish descriptors and obtain insights for the design of amorphous halide electrolytes.

## 2 Computational details

### 2.1 Amorphous structure generation

Initial configurations were constructed with the Packmol package[28] by randomly packing predefined $Li^+$ and $MCl_x^{(x-3)-}$ ionic units, derived from isolated $MCl_x$ cluster geometries taken from crystallographic data,[29,30] into a cubic simulation cell (Fig. 1a). Each composition is the charge-neutral combination $Li_{x-3}MCl_x$ of one $MCl_x^{(x-3)-}$ unit with (x − 3) $Li^+$, where x is the M–Cl coordination number of the stable $MCl_3$ crystal structure in the Materials Project:[29,30] six-coordinate $MCl_6$ units (M = Sc, Y, In, Sm, Ho, Er, Tm, Lu) give $Li_3MCl_6$, eight-coordinate $MCl_8$ units (Tb, Dy, Yb) give $Li_5MCl_8$, and nine-coordinate $MCl_9$ units (La, Pr, Nd, Gd, Ce) give $Li_6MCl_9$ (Table S1). The compositions can be realized experimentally by ball-milling $a MCl_3 \cdot b LiCl$ with $a/b = 1/(x - 3)$. The number of units was set by a target mixture density computed from the tabulated densities of metal chlorides and LiCl, so that the initial cells reflect physically realistic densities prior to equilibration. A 15 × 15 × 15 Å cell (~120 atoms) was used for the DFT benchmark (Section 2.2), and a larger 20 × 20 × 20 Å cell (~280 atoms) for the production simulations to reduce finite-size effects. Periodic boundary conditions were applied in all directions. A Nosé–Hoover thermostat[31–33] and a 1 fs timestep were employed in all MD simulations.

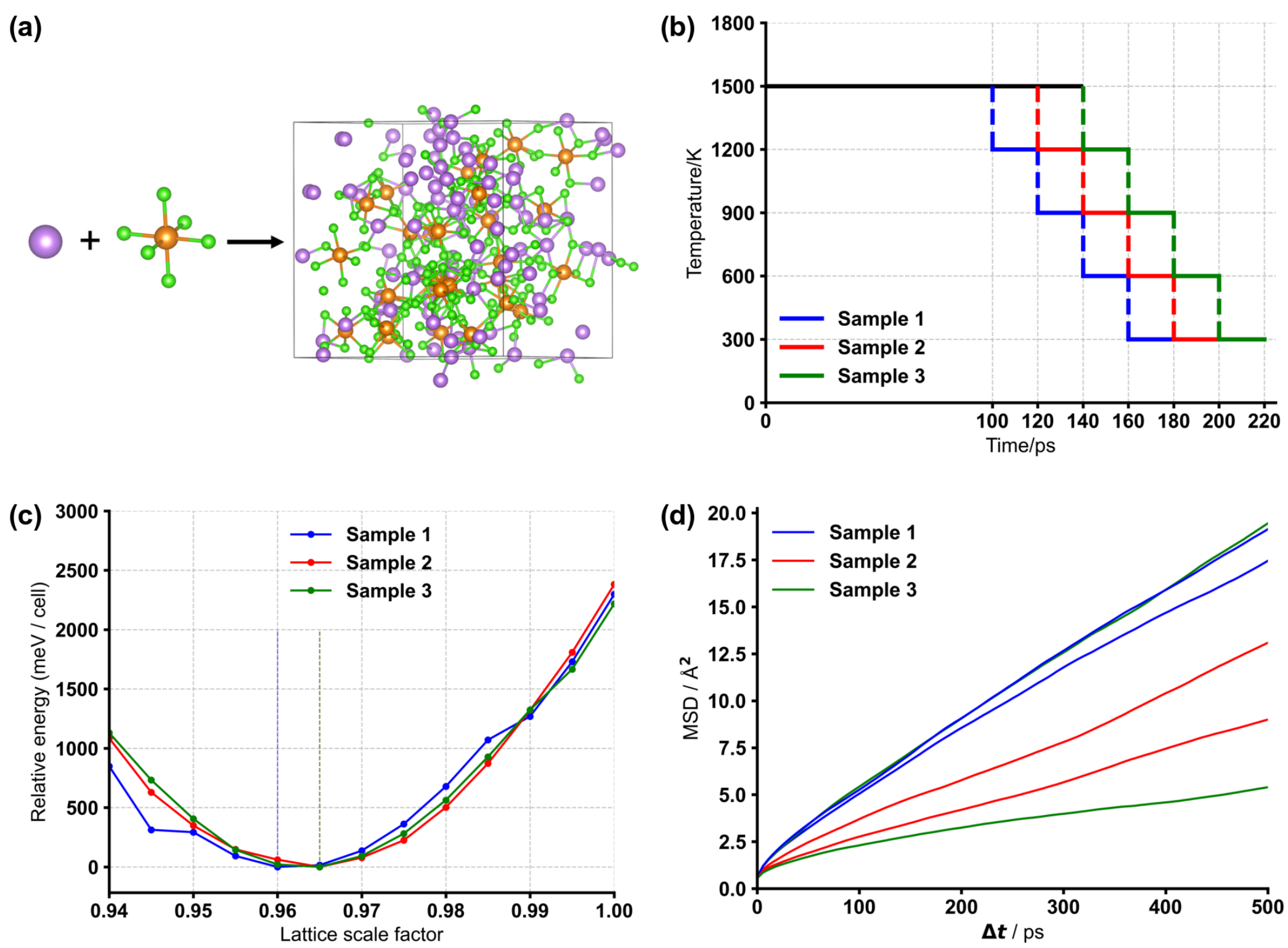

**Fig. 1** Workflow for generating and analyzing amorphous Li–M(III)–Cl materials: (a) Initial configuration built with Packmol from $Li^+$ ions (purple) and $MCl_x$ units (M, orange; Cl, green); (b) Melt sampling and stepwise quenching from 1500 K to 300 K; (c) Relative potential energy versus uniform lattice scale factor for the three quenched configurations; (d) $Li^+$ mean-squared displacement (MSD) versus lag time at 300 K for six trajectories, started from two independent velocity initializations for each of the three quenched structures.

For benchmark calculations, we generated amorphous structures for eighteen materials, which cover $LiTaCl_6$, $Li_3InCl_6$, and sixteen rare-earth chlorides excluding the radioactive promethium. For each composition, twenty snapshots extracted at 1 ps intervals from the 1500 K melt were used in the benchmark. All eighteen materials passed the benchmark (Section 2.2). For these compositions, a single trajectory was equilibrated at 1500 K for 100 ps, and three snapshots drawn from it at 20 ps intervals were quenched independently. Each snapshot was cooled stepwise from 1500 K to 300 K in 300 K decrements, with 20 ps of NVT equilibration at each temperature, followed by a further 20 ps equilibration at 300 K (Fig. 1b). To determine the equilibrium density, the lattice vectors of the quenched cell were uniformly scaled by factors around 1.0 in steps of 0.005; at each scale the internal coordinates were relaxed until the maximum force fell below 0.02 eV Å$^{-1}$, and the lowest-energy configuration was retained (Fig. 1c and Table S1). Production trajectories of 1 ns were then run in the NVT ensemble at 300 K. For the Arrhenius analysis this protocol was repeated at 250, 350, and 400 K. At each temperature, at least six independent 1 ns trajectories were simulated, obtained from the three quenched structures with independent Maxwell–Boltzmann velocity initializations; the error bars in Figs. 2 and 3 are standard deviations across these replicates (Fig. 1d). All screened materials remain amorphous over the 1 ns production runs at 300 K: the M–Cl, Li–Cl, and Li–Li radial distribution functions show no long-range order (Fig. S8), and the $Li^+$ MSD remained linear in every trajectory.

**2.2 DFT benchmarking of the UMA machine-learning interatomic potential**

To validate the universal MLFF UMA (uma-s-1p1),[26] augmented with Grimme's D3 dispersion correction,[34] we benchmarked its energies and forces against density-functional theory (PBE+D3) on 20 snapshots extracted at 1 ps intervals from the 1500 K melt. The dispersion correction is essential: without it, the predicted densities of the amorphous phases fall 20–30% below the reference values, whereas the UMA+D3 model reproduces physically reasonable densities across all compositions. All UMA+D3 simulations were carried out using the Atomic Simulation Environment (ASE).[35] DFT calculations were performed with VASP[36–38] using the Perdew–Burke–Ernzerhof (PBE) exchange–correlation functional[39] with the D3 dispersion correction,[34] based on the projector augmented-wave (PAW) method.[40] A plane-wave cutoff of 520 eV and Γ-point-only Brillouin-zone sampling were used for all calculations. Gaussian smearing (width 0.05 eV) was applied, and the electronic energy was converged to $1 \times 10^{-5}$ eV. Spin-polarized calculations were used for all systems. Static single-point energies and forces were then compared directly with the UMA+D3 predictions (Figs. S1–S3 and S6). We find that the UMA+D3 model exhibits excellent accuracy for all materials in the benchmark, with the energy mean absolute error (MAE) below 1 meV per atom and the force root-mean-square error (RMSE) of 8–25 meV Å$^{-1}$. We exclude $Li_6EuCl_9$ from the screening because its trajectories show spontaneous $Cl_2$ formation at 300 K, which we attribute to plain PBE (UMA's training data[41]) over-stabilizing the reduction of $Eu^{3+}$ ($4f^6$) toward the half-filled $4f^7$ configuration; this pathway is unavailable to the other cations, whose 4f electrons are frozen in the core, absent, or resist reduction. In total, seventeen materials were simulated at 300 K including sixteen Li–M(III)–Cl materials (Fig. 2) and $LiTaCl_6$ (Fig. S5). The energies of amorphous structures are compared with LiCl and $MCl_x$, as well as known crystal phases (Fig. S10). Screened materials are 27 to 45 meV atom$^{-1}$ above their constituent crystal structures, similar to the synthesized $LiTaCl_6$, which is 25 meV atom$^{-1}$ above LiCl + $TaCl_5$ energy, suggesting synthetic feasibility.

We further validate our approach with the prototypical $LiTaCl_6$ system, which has been well-studied both experimentally and computationally. The UMA+D3 model accurately predicts the energies and forces (Fig. S3). It reproduces the radial distribution functions (Fig. S4), Li diffusivity (Fig. S5a), and activation energy (Fig. S5b) compared with the previous study employing VASP MLFF.[22] The calculated activation energy (0.162 ± 0.013 eV) is in excellent agreement with the experimental value of 0.165 eV.[16]

**2.3 Transport properties**

$Li^+$ diffusion coefficients were obtained from the slope of the mean squared displacement (MSD) in its linear regime via the Einstein relation, using the pymatgen package.[42,43] Activation energies were extracted by fitting the temperature dependence of the diffusivity to the Arrhenius relation $D = D_0\exp(-E_a/k_BT)$, using linear fits of $\log_{10}D$ versus 1/T to the mean diffusivities; uncertainties are the standard errors of the fitted slopes. Ionic conductivities were first estimated from the Nernst–Einstein relation; because this approximation neglects ion–ion correlations and is known to overestimate the conductivity of halide electrolytes by a factor of two or more,[18,22,44] the values were rescaled using a single factor defined by the ratio of the experimentally measured conductivity of $LiTaCl_6$ to its simulated $D\cdot\rho$ product (Fig. 4a).

**2.4 Vibrational and cooperative-motion analysis**

The $Li^+$ vibrational density of states (VDOS) was computed as the Fourier transform of the $Li^+$ velocity autocorrelation function from the 300 K production trajectories using the VASPKIT package,[45] and its band center was defined as $\omega_c = \int\omega\cdot\mathrm{VDOS}(\omega)d\omega \,/\, \int\mathrm{VDOS}(\omega)d\omega$. Cooperative $Li^+$ dynamics were quantified by a string analysis:[46] a pair of $Li^+$ ions is deemed correlated within a time interval $\Delta t$ when one ion moves into a position vacated by the other within a cutoff distance $r_s$; correlated pairs sharing an ion are linked into strings, and the number of ions in a string defines the number of correlated bodies N. Ions displacing by more than 3.0 Å within $\Delta t$ that do not belong to any string are counted as independent hops (N = 1). We used $\Delta t$ = 200 ps and $r_s$ = 1.5 Å following a previous study on amorphous halide Li superionic conductors.[23] The sensitivity of $N_{avg}$ to $\Delta t$ is examined in Fig. S9. $N_{avg}$ increases with $\Delta t$ and levels off beyond about 200 ps for nearly all compositions (only $Li_3SmCl_6$ and $Li_3InCl_6$ rise by a further $\leq 0.1$), and the separation between the strongly correlated $Li_6MCl_9$ phases and the weakly correlated $Li_3ScCl_6$ and $Li_3ErCl_6$ is preserved at every window (Fig. S9); $\Delta t = 200$ ps therefore samples the essentially converged string statistics.

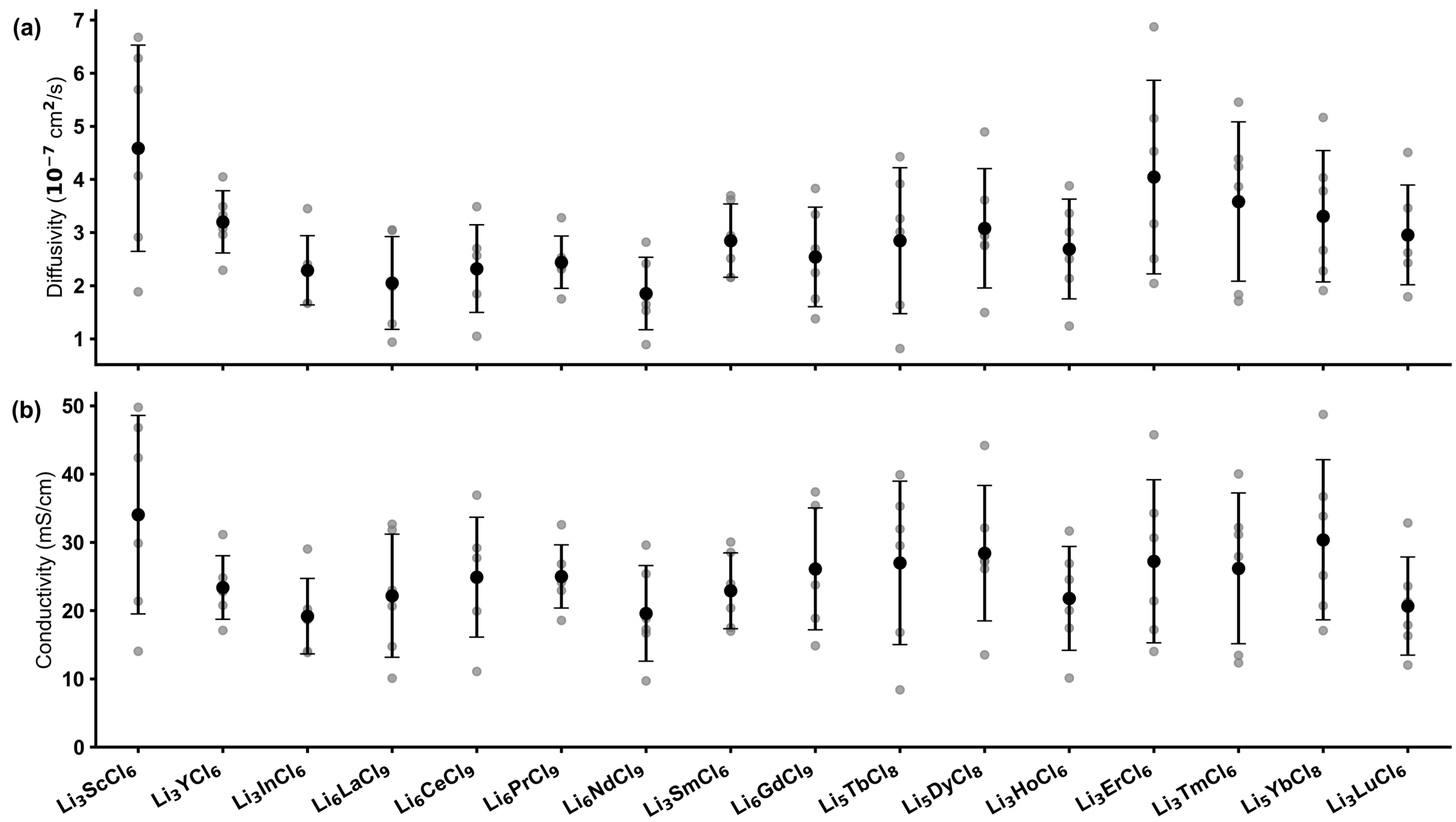

**Fig. 2** Li-ion transport properties of amorphous Li–M(III)–Cl compositions from 1 ns NVT MD simulations at 300 K: (a) $Li^+$ diffusivity; (b) ionic conductivity from the Nernst–Einstein equation. Gray points denote individual values from independent trajectories, and black markers indicate the mean with error bars representing one standard deviation across replicates.

## 3 Results and discussion

### 3.1 Diffusivities, conductivities and activation energies

Molecular dynamics simulations at 300 K reveal substantial $Li^+$ mobility in all amorphous Li–M(III)–Cl compositions. The $Li^+$ diffusivities span roughly 1.8–4.6 × $10^{-7}$ $cm^2$ $s^{-1}$ (Fig. 2a), confirming that fast room-temperature $Li^+$ transport is a general feature of this amorphous halide family rather than a peculiarity of any single composition. Amorphous $Li_3ScCl_6$ exhibits the highest diffusivity (~4.6 × $10^{-7}$ $cm^2$ $s^{-1}$), followed closely by $Li_3ErCl_6$ and $Li_3TmCl_6$, while the chloride-rich $Li_6NdCl_9$ and $Li_6LaCl_9$ lie at the lower end. Notably, the $Li_3MCl_6$ compositions tend to outperform the more lithium-rich $Li_6MCl_9$ phases, indicating that the nominal $Li^+$ concentration alone does not govern performance.

The corresponding ionic conductivities estimated from the Nernst–Einstein equation range from 20 to 34 mS cm$^{-1}$ (Fig. 2b). These values, however, overestimate the true conductivity owing to the neglect of correlated ion motion. Rescaling against the experimental $LiTaCl_6$ reference (Fig. 4a) reduces the conductivities by a factor of about 2.4, placing amorphous $Li_3ScCl_6$ at an estimated 14 mS cm$^{-1}$. Even with this correction and allowing for the systematic uncertainty of the single-point calibration, the estimated conductivities remain competitive with the best halide electrolytes.

Arrhenius analysis over 250–400 K confirms thermally activated transport with excellent linearity (Fig. 3): both representative compositions, $Li_3ScCl_6$ and $Li_5TbCl_8$, yield activation energies of 0.182 ± 0.018 eV ($R^2$ = 0.98) and 0.182 ± 0.010 eV ($R^2$ = 0.99). This low barrier is comparable to those of the benchmark amorphous halides $LiTaCl_6$ (0.165 eV) and $LiNbCl_6$ (0.154 eV). Placing these results in the broader landscape of halide conductors (Fig. 4b), amorphous $Li_3ScCl_6$ ranks among the most conductive amorphous halides, combining a high conductivity with a low activation energy, and clearly outperforms crystalline halide conductors, including $Li_3ScCl_6$, $Li_3InCl_6$, and $Li_2ZrCl_6$. This contrast underscores the benefit of amorphization for $Li^+$ transport in halide electrolytes.

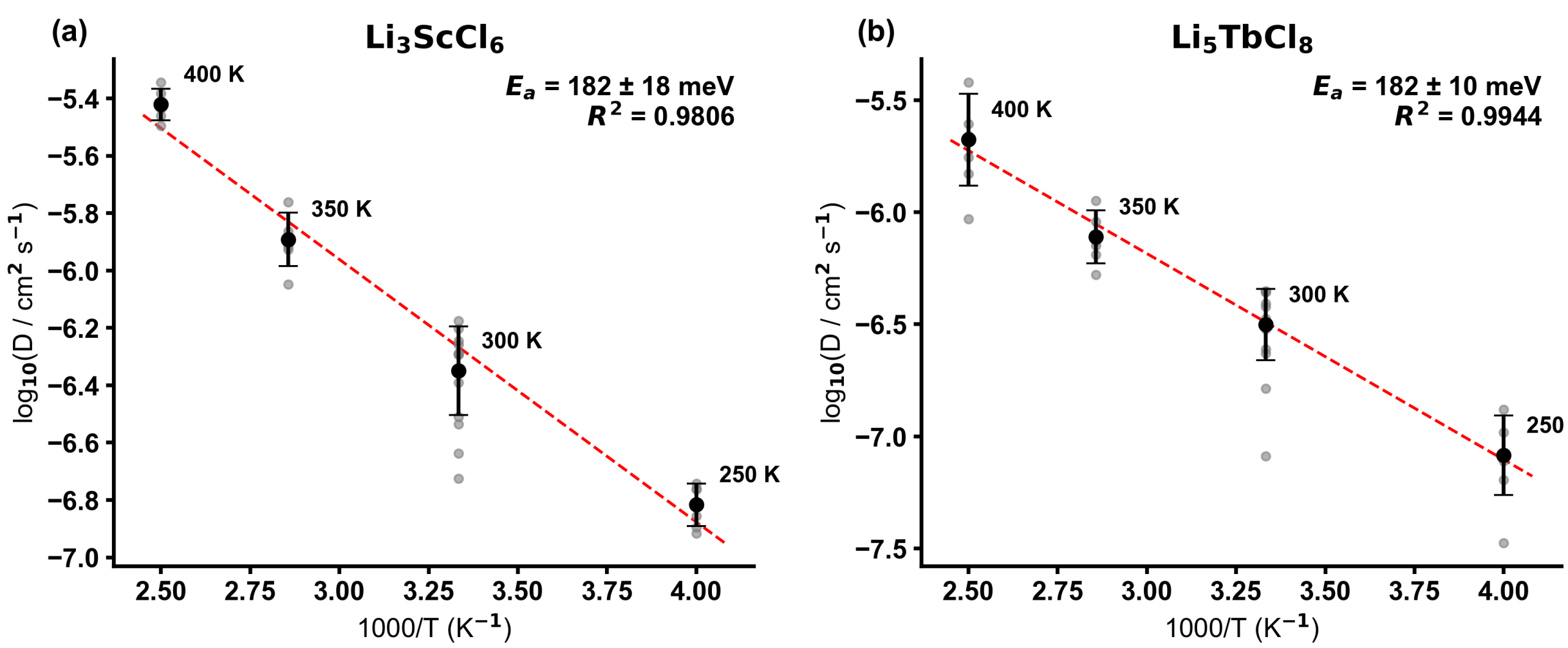

**Fig. 3** Arrhenius analysis of $Li^+$ diffusivity in two representative amorphous Li–M(III)–Cl compositions: (a) $Li_3ScCl_6$; (b) $Li_5TbCl_8$. Gray points are individual trajectories, black markers are means ± one standard deviation, and red dashed lines are linear fits to the means.

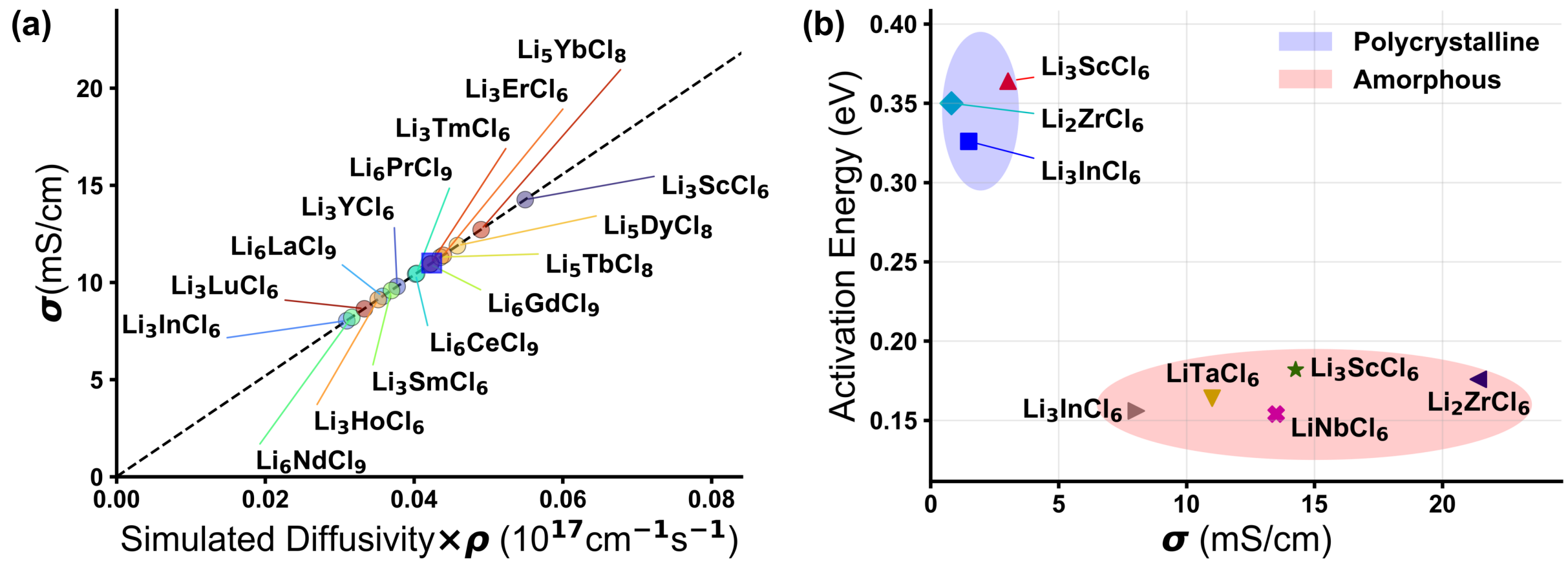

**Fig. 4** (a) Rescaled ionic conductivities of Li–M(III)–Cl compositions, using the linear relation (dashed line) between the experimental $LiTaCl_6$ conductivity (square on the line) and the simulated D·ρ product where ρ is Li-ion concentration. (b) Activation energy versus room-temperature ionic conductivity for representative halide lithium-ion conductors.[7,8,16,18,23,47] Activation energies and room-temperature conductivities were calculated in this work for amorphous $Li_3ScCl_6$ and $Li_3InCl_6$ (Fig. 3a and Fig. S7).

### 3.2 Li diffusivity and VDOS center

To understand the composition dependence of $Li^+$ diffusivity, we turn to the lattice dynamics encoded in the $Li^+$ vibrational density of states (VDOS).[48] The VDOS reflects the stiffness of the local $Li^+$ environment: softer coordination cages shift vibrational weight to lower frequencies and lower the VDOS band center. Fig. 5a compares the normalized $Li^+$ VDOS of three representative compositions; relative to $Li_3YCl_6$ and $Li_6LaCl_9$, the spectrum of $Li_3ScCl_6$ is visibly red-shifted, with reduced spectral weight at high frequencies (above ~10 THz).

Across all screened compositions, the $Li^+$ diffusivity correlates inversely with the Li-VDOS center (Fig. 5b): compositions with lower band centers diffuse faster, with $Li_3ScCl_6$ at the favorable extreme and the chloride-rich $Li_6RECl_9$ phases at the opposite end. This trend is consistent with the soft-lattice picture of ion transport, in which low-frequency, anharmonic vibrations of the coordination shell enlarge the amplitude of $Li^+$ motion and increase the hopping probability. The Li-VDOS center thus serves as a compact, physically transparent descriptor for ranking $Li^+$ diffusivity in amorphous halides.

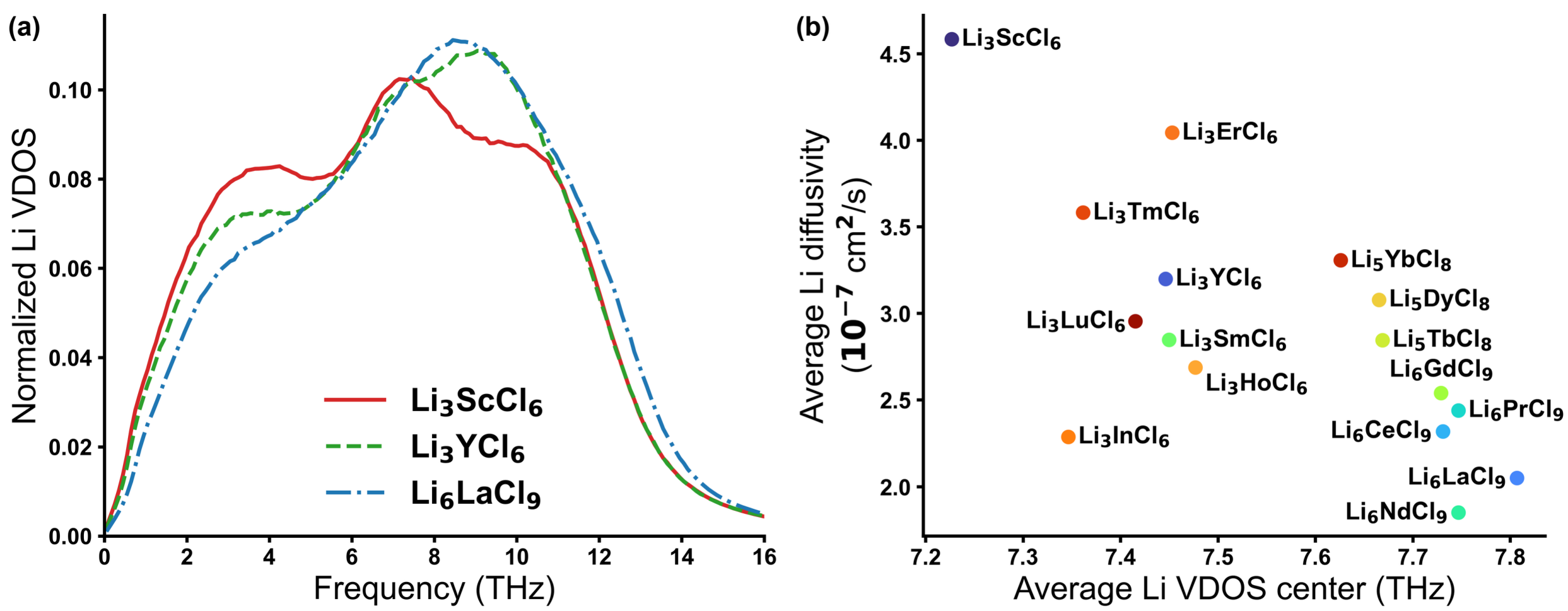

**Fig. 5** (a) Normalized $Li^+$ vibrational density of states (VDOS) for three representative compositions, showing a red shift for $Li_3ScCl_6$ relative to the others. (b) Average $Li^+$ diffusivity at 300 K versus the Li-VDOS center across all screened compositions, revealing an inverse correlation in which lower VDOS centers correspond to higher diffusivities, with $Li_3ScCl_6$ at the favorable extreme.

### 3.3 Li diffusivity and concerted motion

Beyond the softness of the local environment, the degree to which $Li^+$ ions move cooperatively offers a second, complementary axis for understanding transport. We quantified cooperativity with a string analysis that classifies each $Li^+$ displacement as either an independent hop (N = 1) or part of a correlated string of N ions. Individual hopping is the single most probable event in every composition except $Li_6LaCl_9$, $Li_6NdCl_9$, and $Li_6CeCl_9$, but the balance between independent and concerted motion varies systematically across the series: the fraction of independent hops rises from 27% in $Li_6LaCl_9$ to 52% in $Li_3ErCl_6$ (Fig. 6a), and the average number of correlated bodies $N_{avg}$ correspondingly falls from 2.39 to 1.75.

Plotting the diffusivity against $N_{avg}$ reveals a notable trend: in the amorphous Li–M(III)–Cl electrolytes studied here, the compositions with the highest $Li^+$ diffusivity tend to be those in which $Li^+$ migrates least cooperatively (Fig. 6b). The diffusivity generally increases as $N_{avg}$ falls, so that $Li_3ScCl_6$ and $Li_3ErCl_6$, which combine a low $N_{avg}$ with a high fraction of independent hops, are also the fastest conductors, whereas the strongly correlated $Li_6LaCl_9$ and $Li_6NdCl_9$ are the slowest. This inverse correlation contrasts with the behavior commonly reported for crystalline superionic conductors, where concerted, string-like migration is thought to lower the effective barrier and accelerate diffusion.[46,49,50] However, weakly concerted migration has also been reported in other amorphous halide electrolytes[23] and in a high-concentration water-in-salt electrolyte.[51]

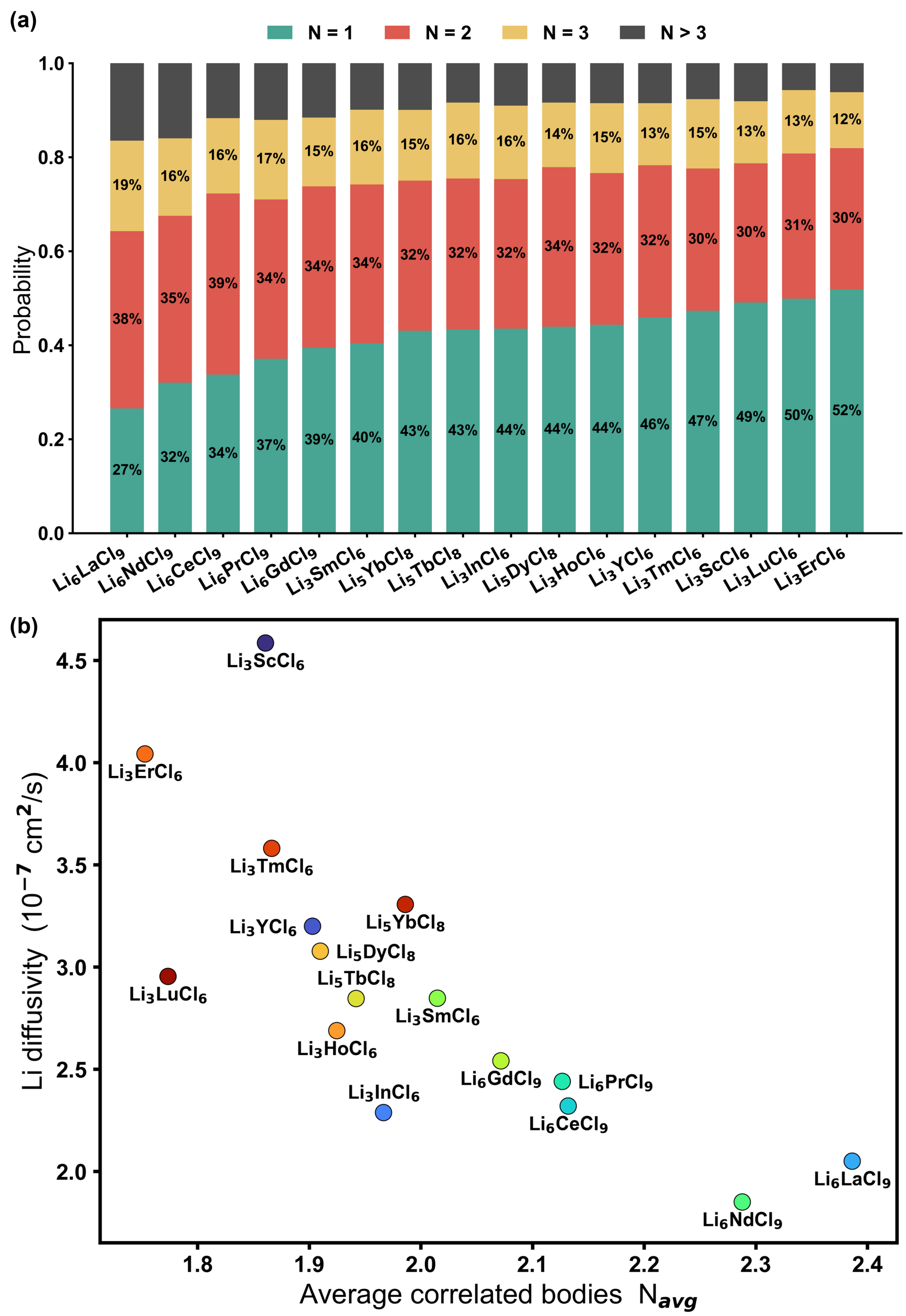

**Fig. 6** Concerted $Li^+$ motion across amorphous Li–M(III)–Cl compositions. (a) Probability distribution of string-like correlated hopping events with N = 1 (independent), 2, 3, and >3 correlated $Li^+$ ions, with compositions ordered by increasing fraction of independent hops. (b)

Average $Li^+$ diffusivity at 300 K versus the average number of correlated bodies $N_{avg}$, showing that compositions dominated by individual hopping (lower $N_{avg}$) tend to exhibit higher $Li^+$ diffusivity.

We attribute this contrast to the fundamental structural difference between crystalline and amorphous conductors. Crystalline lattices provide well-defined, periodically connected channels in which a displaced ion is smoothly replaced by a neighbor, so cooperative strings propagate easily and reduce the migration barrier. The disordered amorphous network lacks such continuous channels; here a high degree of correlation instead signals that a $Li^+$ ion can advance only when several neighbors rearrange in concert — a kinetic bottleneck — whereas in the fastest compositions the abundant, loosely bound interstitial sites permit frequent, low-barrier independent hops. The two descriptors are mutually consistent: the soft, decoupled $Li^+$ environments that lower the VDOS center are precisely those that free individual ions to hop without waiting for collective rearrangement. Taken together, these results generalize across sixteen compositions the weakly concerted migration previously observed in amorphous $Li_2ZrCl_6$ and $Li_2HfCl_6$[23] and suggest a design principle specific to amorphous halides: promote soft, weakly correlated $Li^+$ environments that favor independent hopping, in contrast to the concerted-motion paradigm used to rationalize fast transport in crystalline electrolytes.

## 4 Conclusions

Using molecular dynamics driven by the universal MLFF UMA+D3 and benchmarked against PBE+D3 density-functional theory, we have screened the room-temperature $Li^+$ transport of sixteen amorphous lithium trivalent-metal chloride (Li–M(III)–Cl) electrolytes. All compositions display fast $Li^+$ diffusion (~$1.8$–$4.6 \times 10^{-7}$ $cm^2$ $s^{-1}$). Amorphous $Li_3ScCl_6$ emerges as the standout candidate, combining the highest diffusivity, an estimated conductivity of 14 mS $cm^{-1}$, and a low activation energy of 0.18 eV, placing it among the most conductive amorphous halides and well ahead of its crystalline counterpart. Two descriptors rationalize the composition-dependent transport. First, the $Li^+$ VDOS center correlates inversely with diffusivity, identifying soft $Li^+$ coordination environments as a marker of fast transport. Second, and more strikingly, the average number of correlated bodies in string-like $Li^+$ motion correlates inversely with diffusivity: the most conductive amorphous compositions are those in which $Li^+$ hops independently rather than cooperatively. This finding extends the weakly concerted migration previously observed in amorphous $Li_2ZrCl_6$ and $Li_2HfCl_6$ to a sixteen-member series, contrasts with the concerted-motion paradigm established for crystalline superionic conductors, and points to an amorphous-specific design principle — favor soft, weakly correlated $Li^+$ environments that enable frequent independent hopping. Realizing the ideal amorphous phase for these materials may prove experimentally challenging; in that case, ball-milling with less reactive inorganic salts might stabilize the amorphous phase through entropic effects.

## Author contributions

X. Liu: investigation, formal analysis, visualization, data curation, writing – original draft. S. Xia: investigation, formal analysis. D. Jiang: conceptualization, supervision, funding acquisition, writing – review & editing.

## Conflicts of interest

There are no conflicts to declare.

## Data availability

The optimized amorphous structures, UMA+D3 molecular dynamics trajectories at 1 ps resolution, benchmark energies and forces, and analysis scripts supporting this study are available at https://doi.org/10.5281/zenodo.23093801. The full resolution trajectories are available upon request to the authors.

## Acknowledgements

This research was funded by the U.S. National Science Foundation (DMR 2552465). This research used resources of the National Energy Research Scientific Computing Center (NERSC), a Department of Energy User Facility using NERSC award BES-ERCAP0036958.

## Notes and references